# Polymeric Liquid Layer Densified by Surface Acoustic Wave

*Tianhao Hou,[1,2] Jingfa Yang,[1,2] Wen Wang,[*,2,3] Jiang Zhao[*,1,2]*

[1] *Beijing National Research Center for Molecular Sciences, Institute of Chemistry, Chinese Academy of Sciences, Beijing 100190, China*

[2] *University of Chinese Academy of Sciences, Beijing 100049, China*

[3] *Institute of Acoustics, Chinese Academy of Sciences, Beijing 100190, China*

ABSTRACT.

With the application of surface acoustic wave (SAW) of 39.5 MHz to a model polymer liquid film, polyisobutylene, deposited on the solid substrates, the liquid film is densified, proved by the decrease of film thickness and the increase of refractive index, measured by ellipsometry. Rotational motion of fluorescent probes doped inside the liquid film, measured by polarization-resolved single molecule fluorescence microscopy, is retarded and the dynamical heterogeneity is reduced. It is demonstrated that the application of SAW of high frequency makes the thin polymeric liquid film densified and more dynamically homogeneous.

The dynamical heterogeneity is generic for polymeric and glass forming liquids.[1-4] It is well-recognized that the relaxation dynamics has a very broad distribution in frequency or relaxation time, covering orders of magnitude.[5-7] The dynamical heterogeneity makes the polymeric and glass forming liquid unique while at the same time, bringing about complexity to their properties. It is desirable to create more homogeneous liquid and from it more homogeneous and stable glass can be made.[8-10] Here, we propose a new method to generate better homogeneity in polymeric liquid by applying high frequency mechanical vibration to the system – by applying high frequency surface acoustic wave to a thin polymeric liquid film, the film is densified with the slower molecular dynamics and more dynamical homogeneity.

The polymer adopted is polyisobutylene (PIB, $M_w$ = 300 kg mol$^{-1}$, $M_w/M_n$ = 1.17) with its thin film (20 nm-thick) spin-casted on a quartz crystal. Surface acoustic wave (SAW) is generated by applying electric voltage to the interdigital transducers (IDTs) on the ST-cut quartz crystal. The IDT is fabricated by lithography, capable of generating surface acoustic wave with a wavelength of 80 μm. The detailed information of IDT is provided in Supporting Information. With the speed of SAW is fixed for each media (3158 m s$^{-1}$ for ST-quartz), its central frequency is therefore determined as 39.5 MHz. The mechanical amplitude of the SAW is changeable (< 1 nm), depending on the voltage applied. The whole system is integrated into a compact device so that the temperature of the sample is controlled and more importantly, and it can fit into other instruments for in-situ measurements – the ellipsometer (M-2000 V, J. A. Woollam) for thickness measurements and the single molecule fluorescence microscope so that the rotational dynamics of tracer fluorescent molecules doped can be measured. N,N'-bis(2,6-diisopropylphenyl)-1,6,7,12-tetraphenoxy-3,4,9,10-perylenedicarboximide, (PDI) was adopted as the probe. While the details of the systems are provided in the Supporting information, a more general description is found in Figure 1.

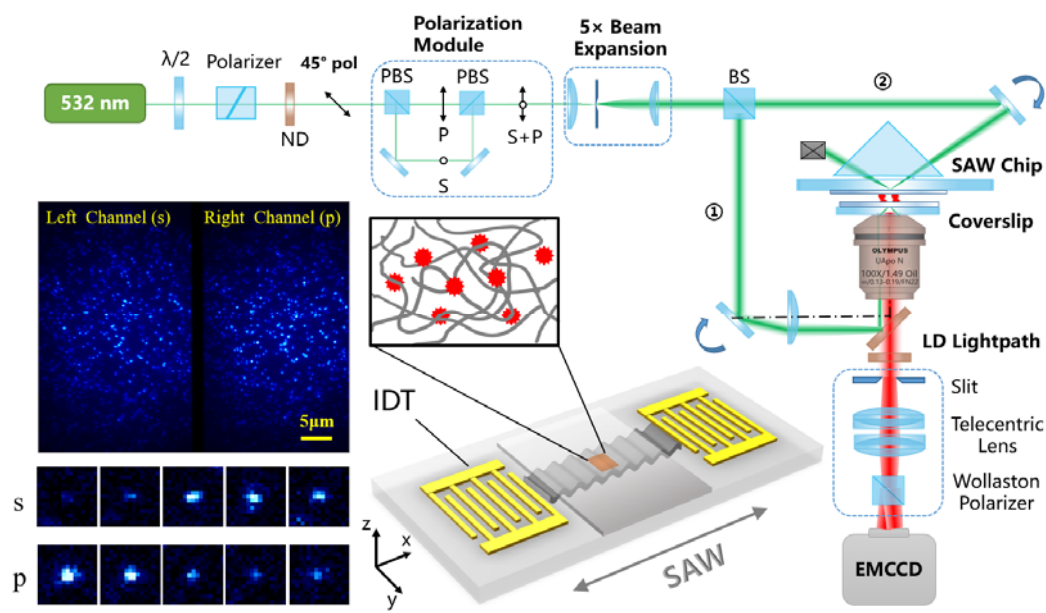

**Figure 1** The schematic of the optical path of the polarization-resolved single molecule fluorescence imaging, and its combination with the SAW device. A typical image of polarization-resolved images of fluorescent molecules acquired by EMCCD is displayed, with the left and right part is the image of two orthogonal polarizations (s and p). A sequence of images of *s* and *p* polarization of one molecule are displayed. The standing wave of SAW generated on the polymer thin film is illustrated between two IDTs.

For single molecule fluorescence microscopy, a focused laser beam comprised of two beams with equal intensity but orthogonal polarizations is introduced to excite the fluorescence in the thin film deposited on a quartz chip. The sample stage is mounted on an inverted microscope (Olympus IX-71) equipped with a high NA objective lens (Olympus UPlanApo 100×, 1.49 N.A.). The excitation is by total internal reflection mode. The fluorescence at the output of the microscope is depolarized by a Wollaston prism so that the image is split into two with orthogonal polarizations and projected onto the imaging chip of an EMCCD camera (Andor 897) separately. In this way, the fluorescence emission images of two orthogonal polarization can be recorded simultaneously and the rotational dynamics of the probes can be analyzed by constructing auto-correlation functions (ACFs) of dichroism of each molecules. [11-12] The images of more than 600 molecules under each condition are analyzed. Two symmetric IDTs are fabricated on the quartz crystal, and by applying AC voltage simultaneously, a standing wave of SAW is created. The images are acquired at more than ten locations of the sample and the area covered are far beyond dimension of the SAW's wavelength.

Figure 2 shows how the thickness and refractive index of the PIB film changes with time under 35 °C, measured by an ellipsometer. It is observed a gradual decrease of the film thickness and an increase of refractive index when the film is at rest, showing the aging and stabilization process. What is emphasized is the behavior of the film when SAW is applied – there occurred a much higher rate of the change. Compared with the case without SAW, where there is 0.2 nm decrease over 466 min, the rate is much higher (film thickness decrease of 0.1 nm within 125 min), showing the acceleration of densification of the PIB driven by SAW.

The rotational dynamics of PDI embedded in PIB thin film with and without the application of SAW is investigated by polarization-resolved single molecule fluorescence imaging method.[11-12] The typical ensemble ACFs of PDI molecules' dichroism with and without the application of SAW are displayed in Figure 3A. The ACFs are fitted with stretched exponential function, $G(\tau) = G(0)\exp\left[-(\tau/\tau_{\text{fit}})^{\beta}\right]$,

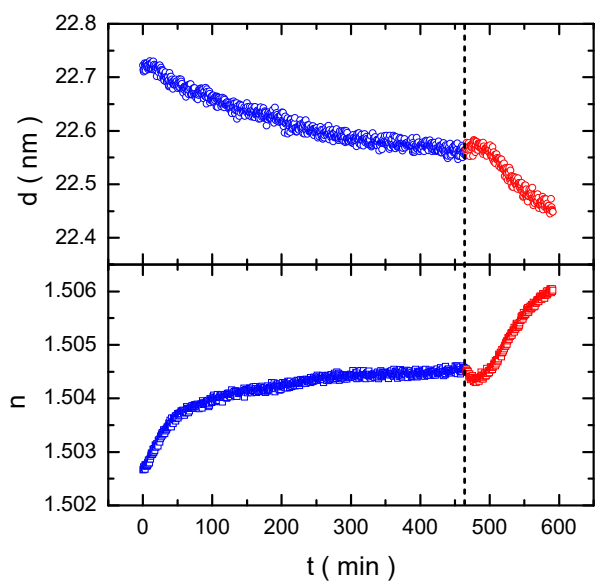

**Figure 2** The value of thickness and refractive index of PIB film as a function of time when the film is annealed at 35 °C. The dash line denotes the moment when SAW is applied.

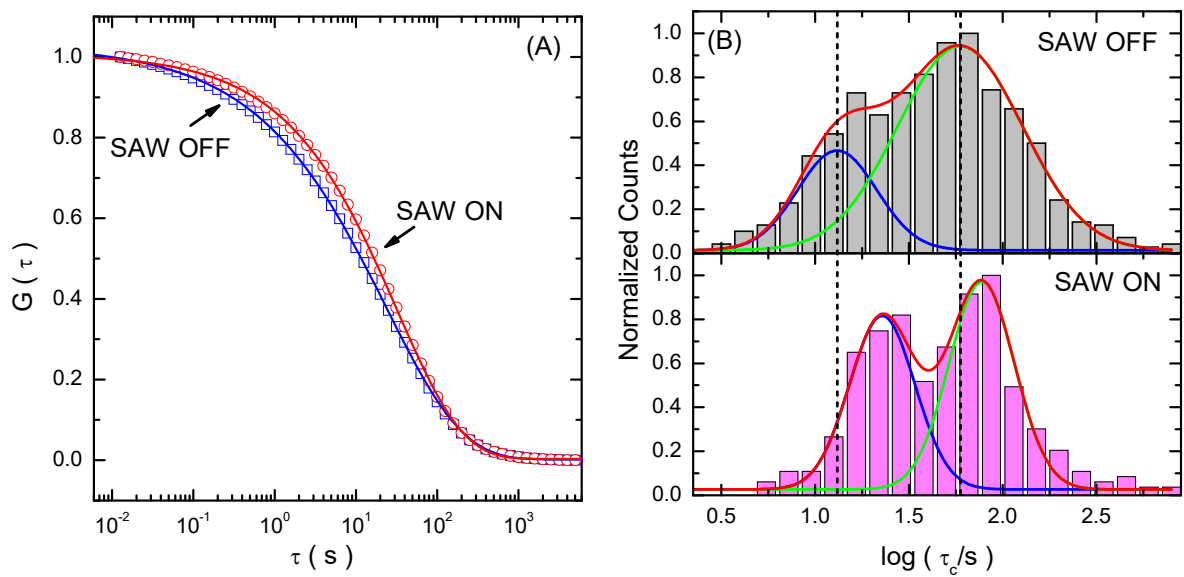

**Figure 3** (A) Typical ensemble auto-correlation of fluorescence dichroism of >600 PDI molecules embedded in PIB thin films, with and without the application of SAW. The data are based on a single-molecule rotation trajectory of 1400 s. (B) The normalized histogram of rotational relaxation correlation time with and without the application of SAW. The film thickness is 20 nm.

and it is discovered that the rotational relaxation time of the case of SAW ON ($\tau_{fit}$ = 32.4 s) is longer than the case without SAW (23.7 s). Meanwhile, the stretched index of SAW ON ($\beta$ = 0.54) is higher than that of SAW OFF (0.46). The results demonstrate that by SAW, the dynamics of the probes inside PIB liquid is retarded while the dynamical heterogeneity is reduced. This finding is further supported by the data of the distributions of characteristic relaxation correlation time ($\tau_c$). The data, as displayed in Figure 3B, are by analyzing ACFs of fluorescence dichroism of individual molecules. Two relaxation modes are recognized, as demonstrated by the well-fitted Gaussian functions. These two modes are attributed to the slower species located closer and the faster one further away to the polymer-solid interface. This is proved by the increase of the relative fraction of faster mode in a thicker film (detailed in Supporting Information).

The dynamics of the probes inside the polymer liquid is retarded by the application of SAW. The values of relaxation correlation time of both modes increase after SAW is applied (Table 1). The FWHM data show the narrowing of the distribution of the relaxation time, indicating the reduction of dynamical heterogeneity. All of the results described above demonstrate that the application of SAW makes the PIB thin liquid film densified and more dynamically homogeneous.

**Table 1. The relaxation time and their distribution of fluorescent probe in PIB films**

| Condition | $\tau_{c1,median}$ (s) | $\tau_{c2,median}$ (s) | FWHM$_1$ | FWHM$_2$ |
| --- | --- | --- | --- | --- |
| SAW OFF | 13.2 | 58.9 | 0.49 | 0.78 |
| SAW ON | 22.9 | 77.6 | 0.41 | 0.41 |

The application of surface acoustic wave with high frequency to the polymer liquid film has created a denser liquid, in which the dynamics of lower frequency is retarded and more dynamically homogeneous. According to the dielectric spectra of PIB (provided in Supporting Information), the frequency of the SAW (39.5 MHz) falls in the high frequency wing of the imaginary part, showing the PIB liquid can absorb the external SAW. The SAW of high frequency excites the motion of very small length scale inside the liquid. The observation indicates the excitation of dynamics of small length scale makes the whole system densified and more dynamically homogeneous, leading to a more stable state of the liquid. It is envisioned that the stable liquid created by SAW can help to make stable glass and further investigation is in process.

# Polymeric Liquid Layer Densified by Surface Acoustic Wave

## Supporting information

*Tianhao Hou, Jingfa Yang, Wen Wang\*, Jiang Zhao\**

**Sample Preparation**

Polyisobutylene (PIB) (Polymer Source) with Mw = 300,000 g·mol$^{-1}$, $M_w/M_n$ = 1.17 were dissolved in toluene (extra-dry spectrophotometric grade, Acros). The substrate, a SAW chip or coverslip, was rinsed with acetone, ethanol and deionized water, respectively, and then treated with oxygen plasma (Harrick Plasma) for 100 s. N,N'-bis(2,6-diisopropylphenyl)-1,6,7,12-tetraphenoxy-3,4,9,10-perylenedicarboximide (PDI) were purchased from Tokyo Chemical Industry (TCI). PDI was dissolved in toluene and further diluted to a concentration of ~8.0 × 10$^{-10}$ M, as measured by fluorescence correlation spectroscopy (FCS).

For film fabrication, the PIB toluene solution of 0.5 wt% was mixed with a tracer amount of PDI (1:100 dilution of fluorescent probe solution to polymer solution). By spin-casting at the rate of 3000 rpm (WS-400-LNPP, Laurell) for 60 s onto a cleaned SAW chip, the liquid film with a thickness of ~20 nm was fabricated. The film thickness was measured by ellipsometry (J.A. Woollam, alpha-SE). Dielectric relaxation spectra of an 80 μm-thick PIB film was measured by dielectric measurement (Novocontrol GMBH Alpha).

For single-molecule fluorescence microscopy measurements, the films were firstly annealed at 110 °C in vacuum for 24 h to remove possible residual solvent and stress. For *in-situ* monitoring by ellipsometry, the films are also annealed at 110 °C in vacuum for 12 h and naturally cooled down to room temperature. Afterwards, the sample is kept in a sample chamber under nitrogen at 35 °C.

**Methodology**

The linear dichroic analysis method can clearly distinct the dynamic heterogeneity with different probe size doped in ~200 nm PVAc ($M_w$ = 500,000 g·mol$^{-1}$, Sigma Aldrich) film at 44 °C, as depicted in **Figure S1**. The measurements yielded results at the same magnitude comparing to Ref. 1.

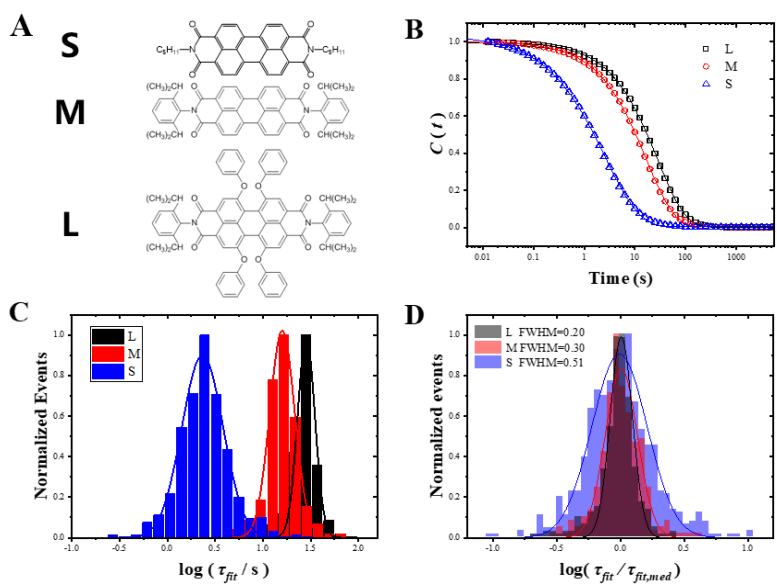

**Figure S1** Results of single molecular linear dichroism analysis with different probes doped in PVAc bulk film at 44 °C. (a) The chemical structures of three perylene diimide derivatives. (b) The ensemble auto-correlation functions (ACF) of each probes. (c) Relaxation time ($\tau_{fit}$) distribution and (d) FWHM comparison of different fluorophores.

Figure S2 shows the schematic of the SAW chip. The electrodes are made of 400 nm-thick aluminum pattern deposition on a ST-cut quartz.

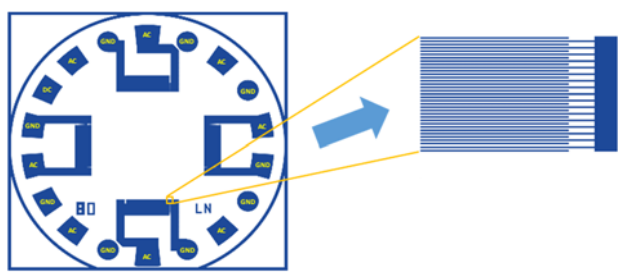

**Figure S2** Schematic of the design of the SAW device, with the right panel showing the detailed structure of the IDT. The finger width is 20 μm and spacing is 20 μm, resulting the repetition period of 80 μm. In experiments, the parallel pair of IDT was used so that a standing wave of SAW is generated at the center of the chip.

Figure S3 shows the designed structure of the device housing the SAW chip and the photos of the parts and ensemble of them, fitted to single molecule fluorescence microscopy with the temperature controller.

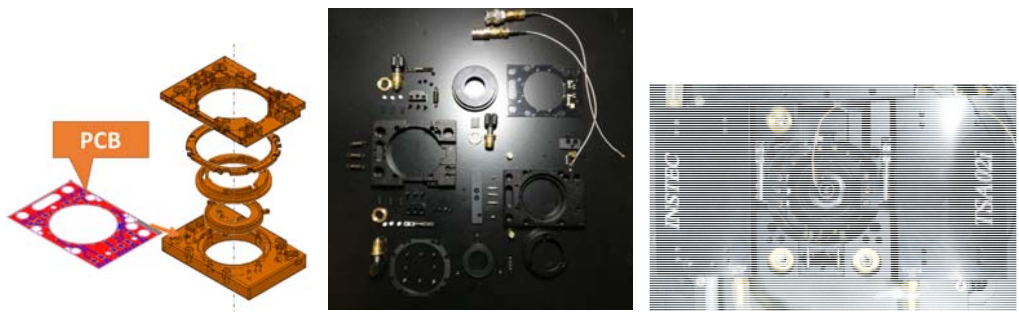

**Figure S3** Schematic and photos of devices housing the SAW chip and fitted to single molecule fluorescence microscopy with temperature control.

**Trajectory length dependence**

For single molecule fluorescence microscopy measurement, photobleaching of the fluorescent probes always leads to trajectories with limited length.[2] As shown in previous single-molecule studies on super-cooled liquids, the exposure time and trajectory length are the factors affecting the results of spatial and temporal heterogeneity.[3-4] Trajectories of different length have been simulated and the dependence of $\tau_c$ and $\beta$ values was investigated and it was found that these parameters reached the stabilized values when the trajectory length is more than $100 \times \tau_c$.[4] In the experiments, measurements were conducted for the maximum length of 1400 s, which is rough 100 times of the relaxation time of the fast mode. **Table S1** demonstrates the comparison of the results with different trajectory length.

Table S1. The comparison of the fitting results for different trajectory length

| Condition | SAW OFF | | | | | | SAW ON | | | | | |
|---|---|---|---|---|---|---|---|---|---|---|---|---|
| Trajectory length | $x_{c1}$ | $x_{c2}$ | FWHM$_1$ | FWHM$_2$ | $A_1$ | $A_2$ | $x_{c1}$ | $x_{c2}$ | FWHM$_1$ | FWHM$_2$ | $A_1$ | $A_2$ |
| 200 s | 0.44 | 0.96 | 0.45 | 0.75 | 0.33 | 0.67 | 0.74 | 1.11 | 0.39 | 0.36 | 0.48 | 0.52 |
| 400 s | 0.66 | 1.27 | 0.45 | 0.67 | 0.26 | 0.74 | 0.94 | 1.41 | 0.55 | 0.43 | 0.53 | 0.47 |
| 1400 s | 1.12 | 1.77 | 0.49 | 0.78 | 0.23 | 0.77 | 1.36 | 1.89 | 0.41 | 0.41 | 0.45 | 0.55 |

Note: $x_{c1}$ and $x_{c2}$ denote the median value of the logarithm of relaxation time of the fast and slow mode ($\log(\tau_c/s)$), respectively. FWHM and $A$ denotes the full width at half magnitude and the fraction,

Quasi-ensemble ACFs, that are the average of all single-molecule ACFs of at least 600 molecules, were calculated for each condition and fitted by stretched exponential function (KWW equation). The fitting results of $\tau_{fit}$ and $\beta$ are displayed in **Table. S2**.

Table S2. Results of fitting of ACF$_{QE}$

| Trajectory length | SAW OFF | | | | SAW ON | | | |
|---|---|---|---|---|---|---|---|---|
| | Number of molecules | $\beta$ | $\tau_{fit}$ | $\tau_c$ | Number of molecules | $\beta$ | $\tau_{fit}$ | $\tau_c$ |
| 200 s | 827 | 0.64 | 4.9 | 6.8 | 884 | 0.86 | 8.4 | 9.1 |
| 400 s | 646 | 0.56 | 9.7 | 16.3 | 736 | 0.67 | 12.9 | 17.2 |
| 1400 s | 624 | 0.46 | 23.7 | 55.4 | 697 | 0.54 | 32.4 | 56.5 |

Note: $\tau_{fit}$ and $\tau_c$ are relaxation time deducted by fitting of the ensemble auto-correlation function and dichroism analysis, respectively.

**Experiments on films of Different Thickness**

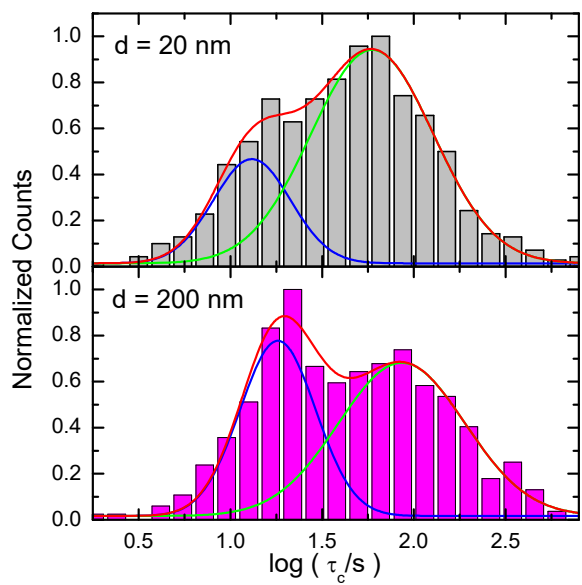

**Figure S4** The normalized histogram of rotational relaxation correlation time inside PIB film of different thickness without the application of SAW. The data are based on trajectory length of 1400 s.

The polarization-resolved single molecule imaging experiments were conducted on PIB films of two thickness (20 and 200 nm) and the data of the distribution of the correlation time of rotational motion are displayed in **Figure S4**.

**Dielectric Relaxation Spectroscopy (DRS)**

The imaginary part of dielectric spectra is shown in **Figure S5A** and the relaxation frequency's dependence on temperature is shown in **Figure S5B**. Data are fitted with the VFTH equation and the result shows the $\tau_{HN}$ value of ~$10^{-6.5}$ s at 293 K, corresponding to a frequency of 3.2 MHz. The SAW frequency (the center frequency of SAW on the bare quart chip is 39.5 MHz, falling in the high frequency wing of the relaxation spectra. The results show that the SAW can be well absorbed by PIB liquid.

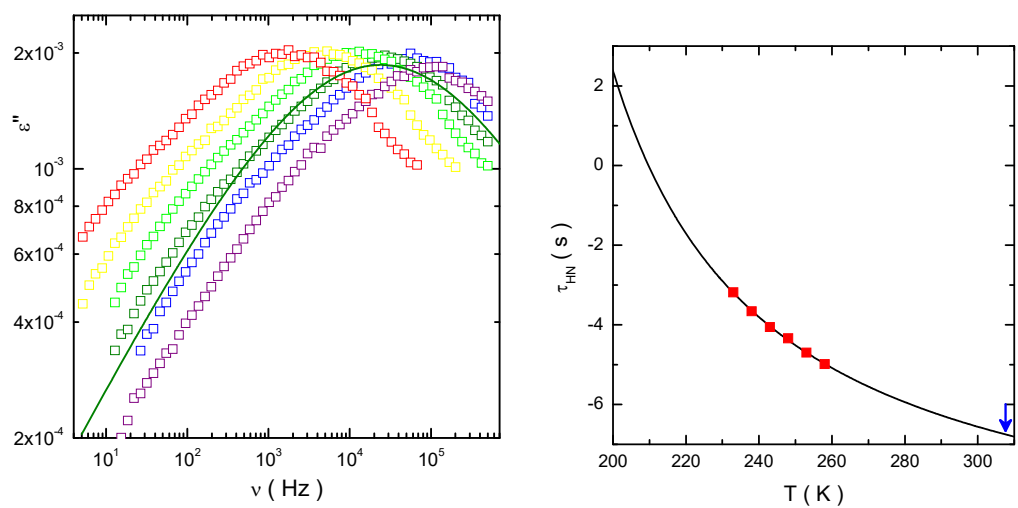

**Figure S5** (A) Dielectric loss spectra of PIB measured from 233 to 258K with the interval of 5 K. Dashed line is the fitting curve of HN (Havriliak-Negami) function. (B) Characteristic time of the α-relaxation, $\tau_{HN}$, as a function of temperature, fitted with VFTH equation. The blue arrow denotes the temperature at which the ellipsometry and single molecule fluorescence microscopy experiments were conducted.